\documentclass[aps,prl,twocolumn,floatfix,superscriptaddress,showpacs]{revtex4}
\usepackage{graphicx}

\begin{document}

\title{Spin-blockade spectroscopy of a two-level artificial molecule}

\author{M. Pioro-Ladri\`{e}re}
\affiliation{Institute for Microstructural Sciences, National Research Council of Canada, Ottawa, Ontario, K1A 0R6 Canada }
\affiliation{Centre de Recherche sur les Propri\'{e}t\'{e}s \'{E}lectroniques de Mat\'{e}riaux Avanc\'{e}s, Universit\'{e} de Sherbrooke, Sherbrooke, Qu\'{e}bec, J1K 2R1 Canada}

\author{M. Ciorga}
\affiliation{Institute for Microstructural Sciences, National Research Council of Canada, Ottawa, Ontario, K1A 0R6 Canada }

\author{J. Lapointe}
\affiliation{Institute for Microstructural Sciences, National Research Council of Canada, Ottawa, Ontario, K1A 0R6 Canada }

\author{P. Zawadzki}
\affiliation{Institute for Microstructural Sciences, National Research Council of Canada, Ottawa, Ontario, K1A 0R6 Canada }

\author{M. Korkusi\'{n}ski}
\affiliation{Institute for Microstructural Sciences, National Research Council of Canada, Ottawa, Ontario, K1A 0R6 Canada }

\author{P. Hawrylak}
\affiliation{Institute for Microstructural Sciences, National Research Council of Canada, Ottawa, 
Ontario, K1A 0R6 Canada }

\author{A. S. Sachrajda}
\affiliation{Institute for Microstructural Sciences, National Research Council of Canada, Ottawa, Ontario, K1A 0R6 Canada }

\date{\today}

\begin{abstract}
Coulomb and spin blockade spectroscopy investigations have been performed on an electrostatically defined ``artificial molecule'' connected to spin polarized leads. The molecule is first effectively reduced to a two-level system by placing both constituent atoms at a specific location of the level spectrum. The spin sensitivity of the conductance enables us to identify the electronic spin-states of the two-level molecule. We find in addition that the magnetic field induces variations in the tunnel coupling between the two atoms. The lateral nature of the device is evoked to explain this behavior.
\end{abstract}


\maketitle

Considerable experimental and theoretical effort is currently being applied towards the realization, characterization and manipulation of artificial molecules \cite{double_dots} building largely on progress in studies of artificial atoms \cite{ciorga00,single_dots,tarucha96}. The main focus is on building spintronic nanostructures, including single and coupled quantum dot devices, in which a single electron or a single spin can be isolated and probed \cite{spintronics}. Connecting a single quantum dot to spin polarized leads effectively resulted in a simple spin transistor \cite{ciorga00,ciorga02,ciorga01,jordan}. Coupled quantum dot systems are technologically important for the creation of entangled spin states for quantum information applications such as the construction of quantum gates \cite{qbits}.      

In this Letter we report on Coulomb and spin blockade spectroscopy studies of a spintronic nanostructure consisting of two laterally coupled quantum dots in series connected to spin polarized leads via tunnel barriers. Although the total number of electrons in each dot was larger than one \cite{comment_1}, it is possible, by applying a magnetic field, to reduce effectively the two-dot system to a two-level molecule, to which we can controllably add  up to four ``valence'' electrons. The above condition is achieved at filling factor $\nu=2$ in the dot, a regime well described in our studies on single dot devices \cite{ciorga02,ciorga01}. In a Fock-Darwin picture of a quantum dot spectrum this corresponds to the region just after the last orbital crossing \cite{tarucha96,tarucha00}. The localized spatial distribution of the relevant wavefunctions allows us to treat electrons occupying the lowest (highest) energy orbitals as ``core'' (valence) electrons. Two valence orbitals, lying at the edge of the individual quantum dots, can be coupled and de-coupled by means of electrostatic gates. The total spin of the valence electrons in the molecular states controls the current through the system, because of the spin polarized leads. The spin of quantum molecules can be tuned in a similar fashion to the spin in single dots---by lowering magnetic field slightly below its $\nu=2$ value and inducing singlet-triplet transitions in the individual dots \cite{ciorga02,tarucha00}. In addition to the the amplitude modulation of the current through the molecule, we also observe variations in the inter-dot tunnel coupling as the magnetic field is reduced. 

Figure 1(a) shows a scanning electron microscope (SEM) image of our coupled quantum dot device. The dots are defined within the two dimensional electron gas (2DEG) of a GaAs-AlGaAs heterostructure. Negative voltages applied to the pairs of gates 1T-0B, 2T-2B and 3T-4B define the lead-to-left-dot, inter-dot and right-dot-to-lead tunnel barriers, respectively. Gates 1B and 3B are used not only to change the number of electrons but also to tune the energy levels of each dot. The dot-to-lead tunnel conductance is always set so that the coupled dot system is well isolated from the leads ($\ll 2e^2/h$), whereas the inter-dot tunnel conductance can be tuned from zero to $2e^2/h$  to control the inter-dot coupling. The device employs the same gate geometry that allowed us to realize single lateral quantum dots containing a single electron \cite{ciorga00,ciorga02}. The conductance $G$ was measured using standard lock-in techniques.

\begin{figure}[ht]
\includegraphics[bb=176 326 421 534,width=0.75\columnwidth,clip]{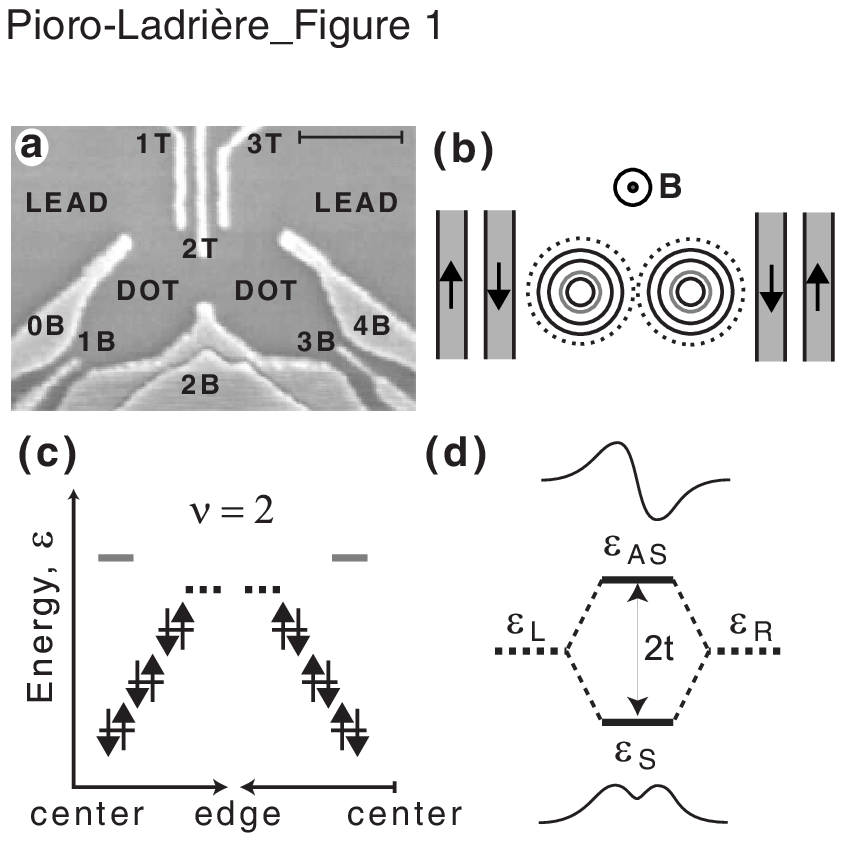}
\label{f1}
\caption{(a) SEM image of the experimental device; scale bar--500 nm.(b) The relevant orbitals in both dots at $\nu=2$ regime. Black (gray) circles indicate initially occupied (unoccupied) orbitals of the 1LL (2LL). Dotted circles represent the outermost empty edge orbitals. The arrows indicate the spin polarization of the leads. (c) Schematic representation of $\nu=2$ ground state in each dot containing an even number of electrons. The horizontal lines correspond to the ring orbitals drawn as circles in (b). (d) Symmetric (S) and anti-symmetric (AS) molecular states formed as a result of the tunnel coupling $t$ between two unoccupied edge orbitals, one from each dot and of equal energy ($\varepsilon_L=\varepsilon_R$).
}
\end{figure}

Consider first two well-isolated dots, each containing an even number of electrons, in the regime of filling factor $\nu=2$. We can picture electrons in this phase as occupying just a simple ladder of states within the first Landau level (1LL), associated with an approximately parabolic confining potential in each dot \cite{nu2}. As depicted in Fig.\ 1(b), the wavefunction of each of these states can be regarded as a ``ring'' with a radius that increases with the energy of the state. Each state is occupied by a pair of electrons with opposite spin. These ``core'' electrons form a spin singlet droplet of electrons within each dot \cite{ciorga02}. We consider these two droplets as the ``vacuum'', state to which ``valence'' electrons are added (Fig.\ 1(c)). 

With both dots set at $\nu=2$, the empty valence orbitals are located at the edges of both droplets. If we now lower the inter-dot barrier, only these outermost valence orbitals become significantly coupled (Fig.\ 1b). The two dots now effectively form a very simple artificial molecule in which valence electrons can occupy two molecular states that are delocalized over both dots \cite{blick}. A representation of the molecular states of the effective two-level system is shown in Fig.\ 1(d) for the case when the energy of the left and the right dot edge orbitals are matched ($\varepsilon_L=\varepsilon_R$). In this case, the molecular states are the related symmetric (S) and anti-symmetric (AS) states. Valence electrons can be manipulated by different operations, eg.  adding/subtracting electrons to/from the system or, at a fixed number of electrons, reducing the magnetic field from its $\nu=2$ value  to transfer an electron between the outermost orbital of the 1LL and the innermost orbital of 2LL within each dot (Fig.\ 1c)\cite{ciorga02}.

We now proceed to demonstrate how Coulomb blockade (CB) spectroscopy is used to measure the energy difference between S and AS molecular states. Figure 2(a) shows a grayscale plot of the measured conductance through the two dots, as a function of voltages $V_L$ and $V_R$ applied to gates 1B and 3B, respectively. The number of core electrons localized in the left and the right dot is 12 and 14, respectively ($N_{core}=(12,14)$). A small mismatch in the occupation number of both dots allows us to separate the effects related to coupling of orbitals belonging to the 1LL and the 2LL, as will be discussed later. The magnetic field is set to $B=1.05$T---in the $\nu=2$ regime. Large Coulomb blockaded regions of zero conductance are separated by two conductance peaks labeled ``A'' and ``B''. The position and amplitude of these peaks vary in the $(V_L,V_R)$ plane. The region where A and B are closest, and their amplitude is largest, corresponds to a pair of triple points, labeled as ``$\alpha$'' and ``$\beta$'', in the charging diagram of the double-dot system \cite{blick}. The CB regions are labeled with numbers corresponding to the number of valence electrons ($N_V=$0, 1 and 2) occupying the two-level system. 

\begin{figure}[t]
\includegraphics[bb=90 501 337 730, width=0.75\columnwidth,clip]{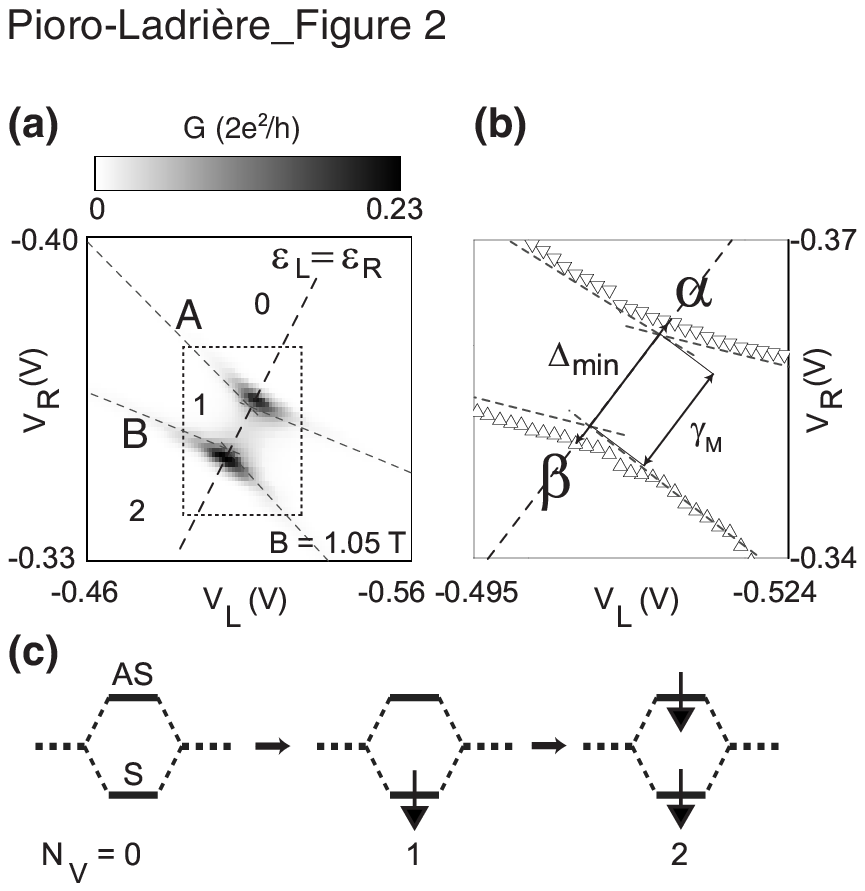}
\label{f2}
\caption{(a) Grayscale plot of the conductance $G$ as a function of the voltages $V_L$ and $V_R$. Two conductance peaks, A and B, separate three Coulomb blockaded regions labeled by the number of valence electrons $N_V$ of the chosen two-level system at $\nu=2$. The faint dashed lines mark the linear portion of the position of both peaks away from the triple points $\alpha$ and $\beta$. (b) Expanded view of the rectangular region in (a) showing the position of both peaks close to $\alpha$ and $\beta$. (c) Filling of the S and AS states for $N_V=$ 0, 1, 2, as deduced from (b).}
\end{figure}

On conductance peak A, at point $\alpha$, electrons tunnel one-by-one from the left lead to the right lead through the S state of the two-level system, and the number of valence electrons fluctuates between 0 and 1 ($N_V=0\leftrightarrow 1$). Peak B corresponds to the situation where $N_V=1\leftrightarrow 2$. The observation that peaks A and B have opposite curvature, i.e. they are not parallel, in Fig.\ 2(b) suggests that at point $\beta$ electrons are transported through the AS state, with the S state permanently occupied with one electron \cite{ziegler}. This simple observation leads us to the level filling scheme depicted in Fig.\ 2(c) which we will confirm by spin blockade spectroscopy discussed below. 

The energy spacing between peaks A and B in Fig.\ 2(b) is given by $\Delta=\gamma_M+\sqrt{(\varepsilon_L-\varepsilon_R)^2+4t^2}$, where $\gamma_M$ is the inter-dot charging energy and $t$  is the tunnel coupling energy \cite{ziegler}. The energies of the left and the right dot edge orbitals, $\varepsilon_L$ and $\varepsilon_R$ , can be adjusted by gate voltages $V_L$ and $V_R$. From the minimum value of the spacing $\Delta$, $\Delta_{min}=\gamma_M+2t$, we estimate $\gamma_M\approx$ 160 $\mu$eV and $t\approx$ 30 $\mu$eV at 1.05 T \cite{blick,conv_factor}. This occurs for the condition $\varepsilon_L=\varepsilon_R$ identified by the black dashed line in the $(V_L,V_R)$ plane which crosses peaks A and B at points $\alpha$ and $\beta$, respectively. When the two levels are detuned by the gates, i.e. $\varepsilon_L\not=\varepsilon_R$, $\Delta$ increases, the S and AS states become bonding and anti-bonding states \cite{pi}, and amplitude of the peaks decreases \cite{blick,ziegler}. 

To investigate the effect of the magnetic field on the spin of the molecular states, in the regime where singlet-triplet transitions are observed in single dots, spin blockade spectroscopy was performed. This spin sensitive technique was originally used for single dot devices \cite{ciorga00,ciorga02,pawel}. Because of the spin polarization of the leads, the tunneling rates for spin-down electrons are higher than for spin-up electrons and the spin injection/detection mechanism is enabled. Hence, if a molecular state is already occupied by a spin-down electron, only the spin-up tunneling channel is available and the current is substantially reduced. Figure 3(a) shows the amplitude of the conductance peaks A and B at points $\alpha$ and $\beta$ ($\varepsilon_L=\varepsilon_R$), related to the addition of valence electrons to the molecular states, as a function of perpendicular magnetic field. Since $N_{core}$ is different for each dot, the position of the triple points (the condition $\varepsilon_L=\varepsilon_R$) in the $(V_L,V_R)$ plane can, and does, change slightly with magnetic field, as seen in Fig.\ 4(a). The origin for this is discussed later. Conductance plots between 0.89 T and 1.05 T, similar to the one shown in Fig.\ 2(a), are carefully repeated and the conductance at points $\alpha$ and $\beta$ extracted. Remarkably, not just the magnetic field dependence of the conductance, but also the amplitude, are practically identical for both $\alpha$ and $\beta$. This reveals that both added electrons must have the same spin. Therefore, due to Pauli exclusion principle, they must be added successively to the S state (at point $\alpha$) and the AS state (point $\beta$). Within the discussed field span we can clearly distinguish three different regimes in terms of the conductance level. These are characterized by ``low''(LO), ``very low'' (VLO) and ``high'' (HI) amplitude. The electronic configurations we deduce for the core electrons for these three regimes are shown in Fig.\ 3(b). The observed variation of the amplitude at points $\alpha$ and $\beta$ will now be explained by a combination of the lateral extent of the S and AS molecular states, depicted in Fig.\ 3(c), and the spin polarized injection/detection mechanism. 

\begin{figure}[t]
\includegraphics[bb=141 229 402 568,clip,width=0.9\columnwidth]{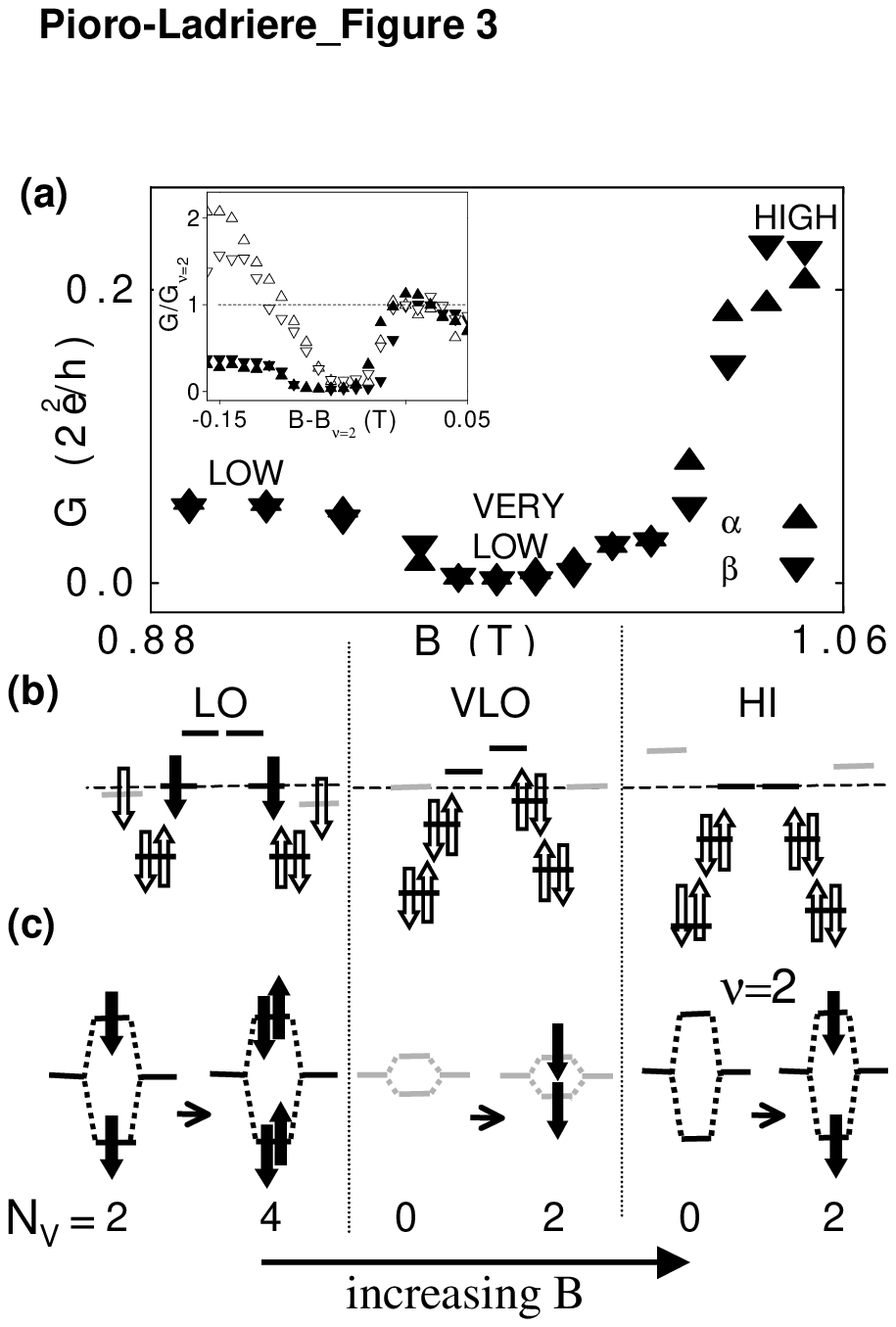}
\label{f3}
\caption{(a)Conductance $G$ at points $\alpha$ and $\beta$ versus perpendicular magnetic field $B$. Inset: $B$ dependence of four consecutive peaks---adding $N_{V}=3,4$ to $N_{core}=(10,12)$ (open symbols) and $N_{V}=1,2$ to $N_{core}=(12,14)$ (solid symbols). All numbers are defined at the $\nu=2$ regime; both $G$ and $B$-positions were normalized to values at the $\nu=2$ transition.(b) Schematic representation of core electrons (open arrows) in each dot in the absence of tunnel coupling, for the three discussed regimes of a magnetic field.  Solid black (gray) lines indicate the orbitals of the 1LL (2LL). (c) Filling of the S and AS states with valence electrons, as deduced from (a).}
\end{figure}

Over the range of magnetic field where the amplitude at points $\alpha$ and $\beta$ is high, electrons move through edge-edge molecular states (originating from the coupling of edge orbitals of each dot) at $\nu=2$ (Fig.\ 1b). As the amplitude at both points is high, it must be a spin-down electron that is added to the system in both $N_V=0\leftrightarrow 1$ and $N_V=1\leftrightarrow 2$ transitions. Hence, during HI phase, for $N_V=1$, one spin-down electron occupies the edge-edge S state, and for $N_V=2$  both edge-edge S and AS states are occupied by one spin-down electron each. Such a level filling scheme, depicted in the rightmost panel of Fig.\ 3(c), confirms the results of CB measurements.

As the magnetic field is decreased away from the $\nu=2$, different orbitals come into play, and the amplitude of points $\alpha$ and $\beta$ falls dramatically as we enter the VLO regime. At fields below about 1.03T the innermost orbital of the 2LL in each dot becomes the lowest unoccupied valence orbital, as illustrated in the middle panel of Fig.\ 3(b) (this is well established from single quantum dot experiments \cite{ciorga02}). As this ``center'' orbital is highly localized at the center of the dot(Fig.\ 1b), it is expected to overlap weakly with the center orbital from the second dot when forming center-center molecular states. This is confirmed when we examine in detail the middle plots of Fig.\ 4. We find that the position of peaks A and B shows almost no curvature near the triple points, i.e. $t\rightarrow 0$ $\mu$eV at 1.00 T \cite{comment_2}. At points $\alpha$ and $\beta$, spin-down electrons may still tunnel through the S and AS states, but a reduction in amplitude at these points is dominated by the reduction in the coupling to the leads. Since the number of core electrons in the two dots is set different, then when the empty edge orbitals in the two dots are aligned in the HI regime (rightmost panel of Fig.\ 3b), the empty center orbital in the left dot lies at a higher energy than the equivalent center orbital in the right dot. For this reason, the $\varepsilon_L=\varepsilon_R$ condition must occur at a slighly different position in the $(V_L,V_R)$ plane for the center-center S and AS states in the VLO regime than for edge-edge states in the HI regime. This is clearly seen in the middle plot of Fig.\ 4(a). Hence, during the VLO phase, for $N_V=1$, one spin-down electron occupies the center-center S state, and for $N_V=2$ both center-center S and AS states are occupied by one spin-down electron each, as depicted in the middle panel of Fig.\ 3(c). 

\begin{figure}[tb]
\includegraphics[bb=134 314 408 521,width=0.85\columnwidth,clip]{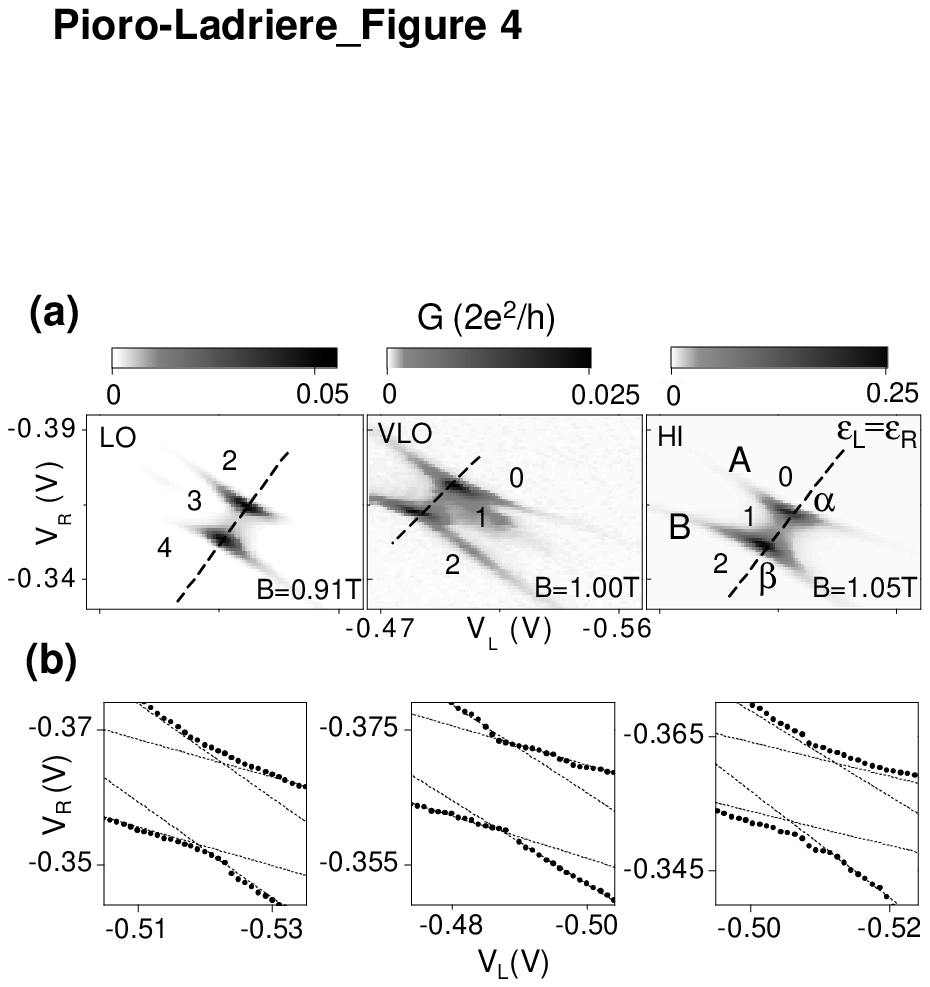}
\label{f4}
\caption{(a) Representative grayscale plots of the conductance in the LO, VLO, and HI regimes. The dashed lines indicate the condition $\varepsilon_L=\varepsilon_R$.(b) Corresponding position of peaks A and B near the triple points $\alpha$ and $\beta$ for each regime. The dashed lines mark the linear portion of the position of both peaks away from the triple points. For details see text.}
\end{figure}

Finally, the LO regime is entered when the magnetic field $B$ is decreased below 0.92 T. The empty center orbital in each dot continues to be lowered in energy relative to the highest occupied edge orbital, so eventually it is energetically favorable for the spin-up electron to transfer from the edge orbital to the center orbital, flipping its spin in the process, as depicted in the leftmost panel of Fig.\ 3(b). This constitutes a singlet-triplet transition for an even-electron dot, well described by studies on single-dot devices \cite{ciorga02,ciorga01,tarucha00}. We see that the triple points move back to a position similar to that in the HI regime, and the tunnel coupling energy recovers to a value close to that observed in the HI regime ($t\approx 25$ $\mu$eV at 0.91 T). From  these observations, it is clear that in the LO regime it is again the edge orbitals that are responsible for the formation of the molecular states, but with the important difference that the edge-edge S and AS states already contain two spin-down electrons (leftmost panel in Fig.\ 3c). Note that the edge valence orbital discussed here is not the one involved in the HI regime but the orbital one-step down in the ladder of states of the 1LL in the respective dots. Lowering $B$ reduces then $N_{core}$ and increases $N_{V}$ by two. Since now initially $N_V=2$ and both the spin-down valence states are full, peaks A and B correspond to filling the edge-edge S and AS states with spin-up electrons \cite{comment_3} and the resulting current is low, due to spin blockade. In a simple picture, the difference between the LO and HI phases is due to the spin blockade of molecular states triggered by the transfer of one electron between the edge and center orbitals of each individual dot. This is supported by observed $B$-dependence of peaks related to adding $N_V=3,4$ at $\nu=2$. The peaks are still paired, but their amplitude pattern is reversed, i.e. HI (LO) becomes LO (HI) while VLO is unchanged. This is shown in the Fig. 3(a) inset.

In summary, we have demonstrated a spintronic nanostructure with an achievable two-level system in which one can determine and manipulate the states of artificial molecules at the single spin level.  The experimental data are fully consistent with our simple model, and lead to a conclusion that spin effects described in this Letter are the artificial molecule analogue of odd-even effects observed for artificial atoms.

\begin{acknowledgments}
We thank D.G. Austing for very helpful discussions. M.P.L. acknowledges financial support from the Natural Sciences and Engineering Research Council of Canada. A.S.S. and P.H. acknowledge the support from the Canadian Institute for Advanced Research.
\end{acknowledgments}



\begin{thebibliography}{}

\bibitem{double_dots}
C. Livermore \textit{et al.}, Science \textbf{274}, 1332 (1996);
T. H. Oosterkamp \textit{et al.}, Nature \textbf{395}, 873 (1998); M. Bayer \textit{et al.}, Science \textbf{291}, 451 (2001); D. S. Duncan \textit{et al.}, Phys. Rev. B \textbf{63}, 45311 (2001); A. W.  Holleitner \textit{et al.}, Science \textbf{297}, 70 (2001).

\bibitem{single_dots}
M. A. Kastner, Phys. Today \textbf{46}, 24 (1993); R. C. Ashoori,  Nature \textbf{379}, 413 (1996); L. Jacak, P. Hawrylak, and A. Wojs, \textit{Quantum Dots} (Springer Verlag, Berlin, 1998).

\bibitem{tarucha96}
S. Tarucha \textit{et al.}, Phys. Rev. Lett. \textbf{77}, 3613 (1996)

\bibitem{ciorga00}
M. Ciorga \textit{et al.}, Phys. Rev. B \textbf{61}, R16315 (2000).

\bibitem{tarucha00}
S. Tarucha \textit{et al.}, Phys. Rev. Lett. \textbf{84}, 2485 (2000).


\bibitem{spintronics}
S. A. Wolf \textit{et al.}, Science \textbf{294}, 1488 (2001). 



\bibitem{ciorga02}
M. Ciorga \textit{et al.}, Phys. Rev. Lett. \textbf{88}, 256804 (2002). 

\bibitem{ciorga01}
M. Ciorga \textit{et al.}, Physica E \textbf{11}, 35 (2001).

\bibitem{jordan}
J. Kyriakidis \textit{et al.}, Phys. Rev. B \textbf{66}, 35320 (2002).

\bibitem{qbits}
J. A. Brum and P. Hawrylak, Superlatt. Microstruct. \textbf{22}, 431 (1997); D. Loss and D. P. DiVincenzo,  Phys. Rev. A \textbf{57}, 120 (1998).


\bibitem{comment_1}
We were able to achieve a one-electron regime while operating the device in a single-dot mode. High level of high frequency telegraph noise, however, made transport measurements of the coupled-dot system, in the regime of one electron per dot, virtually impossible. 
 


\bibitem{nu2}
A. Wensauer, M. Korkusinski, and P. Hawrylak, Phys. Rev B \textbf{67}, 35325 (2003). 

\bibitem{blick}
R. H. Blick \textit{et al.}, Phys. Rev. Lett. \textbf{80}, 4032 (1998).

\bibitem{pi}
M Pi \textit{et al.}, Phys. Rev. Lett. \textbf{87}, 66801 (2001).

\bibitem{ziegler}
R. Ziegler, C. Bruder, and H. Schoeller,  Phys. Rev. B \textbf{62}, 1961 (2000). 


\bibitem{conv_factor}
The value of a conversion factor between $V_{L},V_{R}$ and energy was obtained from high bias voltage experiments.


\bibitem{pawel}
P. Hawrylak \textit{et al.}, Phys. Rev. B \textbf{59}, 2801 (1999).

\bibitem{comment_2}
The accuracy of extracting the tunnel coupling $t$ from our data ($\sim$few $\mu$eV) allows us only to confirm that $t$ is ``close'' to zero, but not if it is ``completely'' switched off.

\bibitem{comment_3}
From the single-dot studies \cite{ciorga02,nu2} we note that for both an odd and even number of electrons only one electron occupies 2LL center orbital of each dot below the $\nu=2$ transition.

\end{thebibliography}
\end{document}